# Fabrication tolerant chalcogenide mid-infrared multimode interference coupler design with applications for Bracewell nulling interferometry

HARRY-DEAN KENCHINGTON GOLDSMITH,[1,*] NICK CVETOJEVIC,[2,3,4] MICHAEL IRELAND[5] AND STEPHEN MADDEN.[1]

[1]*Centre for Ultrahigh bandwidth Devices for Optical Systems (CUDOS), Laser Physics Centre, Research School of Physics and Engineering, Australian National University, ACT 2601, Australia*
[2]*The Australian Astronomical Observatory, level 1, 105 Delhi Rd, North Ryde, NSW 1670, Australia*
[3]*Institute of Photonics & Optical Science, School of Physics, University of Sydney, NSW 2006, Australia*
[4]*Centre for Ultrahigh bandwidth Devices for Optical Systems (CUDOS), School of Physics, University of Sydney, NSW 2006, Australia*
[5]*Research School of Astronomy & Astrophysics, Australian National University, Canberra, ACT 2611, Australia*
*\*harry-dean.kenchingtongoldsmith@anu.edu.au*

**Abstract:** Understanding exoplanet formation and finding potentially habitable exoplanets is vital to an enhanced understanding of the universe. The use of nulling interferometry to strongly attenuate the central star's light provides the opportunity to see objects closer to the star than ever before. Given that exoplanets are usually warm, the 4 µm Mid-Infrared region is advantageous for such observations. The key performance parameters for a nulling interferometer are the extinction ratio it can attain and how well that is maintained across the operational bandwidth. Both parameters depend on the design and fabrication accuracy of the subcomponents and their wavelength dependence. Via detailed simulation it is shown in this paper that a planar chalcogenide photonic chip, consisting of three highly fabrication tolerant multimode interference couplers, can exceed an extinction ratio of 60 dB in double nulling operation and up to 40 dB for a single nulling operation across a wavelength window of 3.9 to 4.2 µm. This provides a beam combiner with sufficient performance, in theory, to image exoplanets.

## References and Links

1. A. L. Kraus and M. J. Ireland, "LkCa 15: A young exoplanet caught at formation?," ApJ **745**, 5 (2012).
2. A. A. Michelson, "Measurement of Jupiter's satellites by interference," Nature **45**, 160-161 (1891).
3. A. A. Michelson and F. G. Pease, "Measurement of the diameter of alpha Orionis with the interferometer," ApJ **53**, 249-259 (1921).
4. A. A. Michelson, "On the Application of interference methods to astronomical measurements," ApJ **51**, 257-262 (1920).
5. J. E. Graves, F. J. Roddier, M. J. Northcott, J. Anuskiewicz, and G. J. Monnet, "Adaptive optics at the University of Hawaii IV: a photon-counting curvature wavefront sensor," Proc. SPIE **2201**, 502-507 (1994).
6. G. Rousset, J.-L. Beuzit, N. N. Hubin, E. Gendron, P.-Y. Madec, C. Boyer, J.-P. Gaffard, J.-C. Richard, M. Vittot, P. Gigan, and P. J. Lena, "Performance and results of the COME-ON+ adaptive optics system at the ESO 3.6-m telescope," Proc. SPIE **2201**, 1088-1098 (1994).
7. N. Madhusudhan, A. Burrows, and T. Currie, "Model atmospheres for massive gas giants with thick clouds: Application to the Hr 8799 planets and predictions for future detections," ApJ **737**, 34, (2011).
8. C. Marois, B. Macintosh, T. Barman, B. Zuckerman, I. Song, J. Patience, D. Lafrenière, and R. Doyon, "Orbiting the Star HR 8799," Science **322**, 1348-1352 (2008).
9. A. Burrows, D. Sudarsky, and I. Hubeny, "Spectra and diagnostics for the direct detection of wide-separation Extrasolar Giant Planets," ApJ **609**, 407-416 (2004).
10. D. A. Fischer, R. P. Butler, and G. W. Marcy, "A sub-Saturn mass planet orbiting HD 3651," ApJ **44**, 3325-3335 (2003).
11. R. N. Bracewell, "Detecting nonsolar planets by spinning infrared interferometer," in *Nature,* 780-781 (1978).
12. O. Guyon, F. Martinache, R. Belikov, and R. Soummer, "High Performance Piaa Coronagraphy With Complex Amplitude Focal Plane Masks," ApJS **190**, 220-232 (2010).

13. N. Jovanovic, P. G. Tuthill, B. Norris, S. Gross, P. Stewart, N. Charles, S. Lacour, M. Ams, J. S. Lawrence, A. Lehmann, C. Niel, J. G. Robertson, G. D. Marshall, M. Ireland, A. Fuerbach, and M. J. Withford, "Starlight demonstration of the Dragonfly instrument: An integrated photonic pupil-remapping interferometer for high-contrast imaging," MNRAS **427**, 806-815 (2012).
14. J.-P. Berger, P. Haguenauer, P. Kern, K. Perraut, F. Malbet, I. Schanen, M. Severi, R. Millan-Gabet, and W. Traub, "Integrated optics for astronomical interferometry," Astronomy and Astrophysics **376**, 163-164 (2001).
15. L. Labadie, J.-P. Berger, N. Cvetojevic, R. Haynes, R. Harris, N. Jovanovic, S. Lacour, G. Martin, S. Minardi, G. Perrin, M. Roth, and R. R. Thomson, "Astronomical photonics in the context of infrared interferometry and high-resolution spectroscopy," Proc. SPIE **9907**, 990718 (2016).
16. G. Li, T. Eckhause, K. A. Winick, J. D. Monnier, and J.-P. Berger, "Integrated optic beam combiners in lithium niobate for stellar interferometer," Proc. SPIE **6268**, 626834, (2006).
17. H.-k. Hsiao, K. a. Winick, J. D. Monnier, and J. P. Berger, "An infrared integrated optic astronomical beam combiner for stellar interferometry at 3-4 microm," Opt. Express **17**, 18489-18500 (2009).
18. H.-k. Hsiao, K. a. Winick, and J. D. Monnier, "Midinfrared broadband achromatic astronomical beam combiner for nulling interferometry," Appl. Opt. **49**, 6675-6688 (2010).
19. J. D. Monnier, M. J. Ireland, S. Kraus, F. Baron, M. Creech-Eakman, R. Dong, A. Isella, A. Merand, E. Michael, S. Minardi, D. Mozurkewich, R. Petrov, S. Rinehart, T. ten Brummelaar, G. Vasisht, E. Wishnow, J. Young, and Z. Zhu, "Architecture design study and technology road map for the Planet Formation Imager (PFI)," Proc. SPIE **9907**, 99071O (2016).
20. M. J. Ireland, J. D. Monnier, S. Kraus, A. Isella, S. Minardi, R. Petrov, T. ten Brummelaar, J. Young, G. Vasisht, D. Mozurkewich, S. Rinehart, E. A. Michael, G. van Belle, and J. Woillez, "Status of the Planet Formation Imager (PFI) concept," Proc. SPIE **9907**, 99071L (2016).
21. B. Norris, N. Cvetojevic, S. Gross, N. Jovanovic, P. N. Stewart, N. Charles, J. S. Lawrence, M. J. Withford, and P. Tuthill, "High-performance 3D waveguide architecture for astronomical pupil-remapping interferometry," Opt. Express **22**, 18335-18353 (2014).
22. S. Seager, "The search for extrasolar Earth-like planets," Earth Planet. Sci. Lett. **208**, 113-124 (2003).
23. F. Selsis, L. Kaltenegger, and J. Paillet, "Terrestrial exoplanets: diversity, habitability and characterization," Phys. Scr. **T130**, 014032 (2008).
24. A. Rogalski, "Organic field-effect transistors," Opto-Electronics Review **18**, 121-136 (2010).
25. F. Li, S. Jackson, E. Magi, C. Grillet, S. Madden, Y. Moghe, A. Read, S. G. Duvall, P. Atanackovic, B. J. Eggleton, and D. J. Moss, "Low propagation loss silicon-on-sapphire integrated waveguides for the mid-infrared," Opt. Express **19**, 15212-15220 (2011).
26. L. Carletti, P. Ma, Y. Yu, B. Luther-Davies, D. Hudson, C. Monat, R. Orobtchouk, S. Madden, D. J. Moss, M. Brun, S. Ortiz, P. Labeye, S. Nicoletti, and C. Grillet, "Nonlinear optical response of low loss silicon germanium waveguides in the mid-infrared," Opt. Express **23**, 8261-8271 (2015).
27. M. Smit, X. Leijtens, H. Ambrosius, E. Bente, J. v. d. Tol, B. Smalbrugge, T. d. Vries, E.-J. Geluk, J. Bolk, R. v. Veldhoven, L. Augustin, P. Thijs, D. D'Agostino, H. Rabbani, K. Lawniczuk, S. Stopinski, S. Tahvili, A. Corradi, E. Kleijn, D. Dzibrou, M. Felicetti, E. Bitincka, V. Moskalenko, J. Zhao, R. Santos, G. Gilardi, W. Yao, K. Williams, P. Stabile, P. Kuindersma, J. Pello, S. Bhat, Y. Jiao, D. Heiss, G. Roelkens, M. Wale, P. Firth, F. Soares, N. Grote, M. Schell, H. Debregeas, M. Achouche, J.-L. Gentner, A. Bakker, T. Korthorst, D. Gallagher, A. Dabbs, A. Melloni, F. Morichetti, D. Melati, A. Wonfor, R. Penty, R. Broeke, B. Musk, and D. Robbins, "An introduction to InP-based generic integration technology," Semicond. Sci. Technol. **29**, 083001 (2014).
28. L. Ottaviano, P. Minhao, E. Semenova, and K. Yvind, "Low-loss high-confinement waveguides and microring resonators in AlGaAs-on-insulator," Opt. Lett. **41**, 3996-3999 (2016).
29. L. E. Myers, W. R. Bosenberg, R. C. Eckardt, M. M. Fejer, and R. L. Byer, "Multigrating quasi-phase-matched optical parametric oscillator in periodically poled LiNbO3," Opt. Lett. **21**, 591-593 (1996).
30. P. Ma, D.-Y. Choi, Y. Yu, X. Gai, Z. Yang, S. Debbarma, S. Madden, and B. Luther-Davies, "Low-loss chalcogenide waveguides for chemical sensing in the mid-infrared," Opt. Express **21**, 29927-29937 (2013).
31. A. Arriola, S. Mukherjee, D. Choudhury, L. Labadie, and R. R. Thomson, "Ultrafast laser inscribed integrated waveguide components for L-band interferometry," Proc. of SPIE **9146**, 91462L (2014).
32. S. Madden, Z. Jin, D. Choi, S. Debbarma, D. Bulla, and B. Luther-Davies, "Low loss coupling to sub-micron thin film deposited rib and nanowire waveguides by vertical tapering," Opt. Express **21**, 3582-3594 (2013).
33. D. a. P. Bulla, R. P. Wang, a. Prasad, a. V. Rode, S. J. Madden, and B. Luther-Davies, "On the properties and stability of thermally evaporated Ge-As-Se thin films," Appl. Phys. A Mater. Sci. Process. **96**, 615-625 (2009).
34. R. Errmann, S. Minardi, L. Labadie, B. Muthusubramanian, F. Dreisow, S. Nolte, and T. Pertsch, "Interferometric nulling of four channels with integrated optics," Appl. Opt. **54**, 7449-7454 (2015).
35. J. R. P. Angel and N. J. Woolf, "An imaging nulling interferometer to study extrasolar planets," ApJ **475**, 373-379 (1997).
36. T. Han, "Nano-Moulding of Integrated Optical Devices," (Australian National University, 2011). Thesis.
37. P. A. Besse, M. Bachmann, H. Melchior, L. B. Soldano, and M. K. Smit, "Optical bandwidth and fabrication tolerances of multimode interference couplers," J. Light. Technol. **12**, 1004-1009 (1994).
38. L. B. Soldano and E. C. M. Pennings, "Optical multi-mode interference devices based on self-imaging: principles and applications," J. Light. Technol. **13**, 615-627 (1995).

39. M. T. Hill, X. J. M. Leijtens, G. D. Khoe, and M. K. Smit, "Optimizing imbalance and loss in 2 X 2 3-dB multimode interference couplers via access waveguide width," J. Light. Technol. **21**, 2305-2313 (2003).
40. O. Bryngdahl, "Image formation using self-imaging techniques," J. Opt. Soc. Am. **63**, 416-419 (1973).
41. M. Bachmann, P. A. Besse, and H. Melchior, "General self-imaging properties in N × N multimode interference couplers including phase relations," Appl. Opt. **33**, 3905-3911 (1994).
42. M. Bachmann, P. A. Besse, and H. Melchior, "Overlapping-image multimode interference couplers with a reduced number of self-images for uniform and nonuniform power splitting," Appl. Opt. **34**, 6898-6910 (1995).

1. **Introduction**

The search for planets in star systems other than our own (exoplanets) is ongoing in the field of astrophysics. Ultimately, the atmospheric characterization of these exoplanets, especially ones considered Earth-like, is vital to understanding how prolific habitable planets are in the universe and ultimately if life, as we know it, exists outside of our solar system. Interferometry is key to future spectral analysis of exoplanets.

Amongst the available methods for direct imaging, interferometry has come to the forefront of exoplanet hunting after the infrared imaging of a planet in its early stages of formation was reported [1]. Interferometry, as an astronomical application, was first used by Michelson to image the moons of Jupiter [2] and other, larger, stellar phenomena [3,4]. Interferometry has the advantage of improved accuracy when measuring the Fourier components of barely resolved objects.

Space based instruments will be the ultimate incarnation for astronomical interferometry because of the distortions imposed by the Earth's atmosphere, however, ground based interferometers exploiting advanced adaptive optics (AO) systems [5,6] have significantly improved terrestrial systems to the point that useful science is now possible even with large apertures.

Exoplanetary light is many orders of magnitude weaker than that of the host star, as it is either a weak reflection of the host star (attenuated by distance and albedo) or comprises low (compared to the host star) temperature thermal emission for young exoplanets. To image planets in such systems clearly requires a means of blocking the host star's light to render the exoplanetary light detectable. As an example, for a planet in the HR 8799 system recently studied with direct imaging, the contrast ranges between 37 and 44 dB [7, 8]. To cite another example providing contrast requirements, consider extrasolar giant planets (EGPs) that derive their luminosity directly from reprocessed starlight. Such planets have contrasts at 4 μm of ~100 dB at the orbit of our Jupiter (5 astronomical units, AU), decreasing to ~40 dB for orbits ~0.2 AU [9]. Such exoplanets have already been identified: for example HD 3651b, at a distance of only 11 pc, was discovered using radial velocity methods [10] and has a semi-major axis of 0.284 AU.

Common exoplanets that reside at a separation of ~1 AU from their hosts are too close to their host star to attain sufficient contrast with commonly used coronagraphic imaging methods where the central star is blocked behind a mask. Either nulling interferometry (first proposed by Bracewell in 1978 [11]) or advanced coronagraphic techniques such as complex phase mask based coronagraphs with pupil apodization to eliminate the telescope diffraction pattern are required (e.g. [12]). Interferometric nullers have several advantages over even advanced coronagraphs; leaving aside the advantages in fabrication and alignment, nullers offer throughput and contrast that are significantly improved at small inner working angles ($\theta \sim \lambda/D$ or smaller), meaning imaging can be performed much closer to the host star.

The advantages of using a nuller come at a price, and as with all broadband interferometers, path matching constraints, environmental stability, and alignment are all serious issues in bulk optics based interferometer implementations. It is clear that a photonic chip based interferometer has much to offer in ameliorating these difficulties. This has been clearly demonstrated in the Near-Infrared J and H band (1.33 µm and 1.55 µm) [13] with the first ever integrated optic device to combine stellar signals achieved in 2001 [14] in the H band. However, achieving the same architecture in the Mid-Infrared (MIR) is rather more challenging.

A recent review into the astronomical context of infrared interferometry [15] highlights the need for nulling interferometers. Designs of various infrared nulling architecture have been made in Lithium Niobate [16,17] and Silicon-Germanium [18] but, to our knowledge, no experimental nulling depths, theoretical projections of expected performance, nor extensive fabrication tolerance analyses incorporating real world inaccuracies have been published for interferometry devices in these materials.

In this paper a critical part of such an interferometer (the beam combiner) is first design optimized using simulations with the beam propagation method (BPM), and the performance across the operational bandwidth is investigated over the expected range of real world fabrication tolerances. Detailed simulations of 3 dB splitters in the form of multimode interference couplers (MMIs) are presented to show that a fabrication tolerant, high performance, wide bandwidth, planar waveguide design for a basic MIR nulling interferometer is possible. Such a device could then be utilized to find exoplanets either in a long baseline interferometer such as the Planet Formation Imager [19, 20] or as the MIR beam combiner for an instrument such as *Dragonfly* [21] on a large ground based telescope.

## 2. Photonics background

The MIR is particularly favourable for exoplanet research as the contrast ratio between star and exoplanet light is maximized at two windows [7], specifically in the L (3 to 4 μm) and N (7.5 to 14 μm) bands [22]. The N band is most interesting for mature exoplanets at Earth-like temperatures, especially at 9.7 μm where the $O_3$ absorption is visible [23]. In the N band, however, the detector performance is typically two orders of magnitude worse compared to the L band [24], thermal emissions from room temperature optics (e.q. the telescope primary) and other environmental emissions decrease signal-to-noise ratios immensely, and diffraction-limited angular resolution is poorer. The L band is especially advantageous for the study of exoplanetary formation as the relatively high temperatures (~700K) of such bodies moves the black body emission peak to the L band region. As exoplanet formation is a hotly debated process, interferometric nullers in the 4 μm wavelength band are of current interest.

To create a monolithic chip at 4 μm, traditional silicate based glasses normally used in planar integrated optics are of no utility as they are no longer transparent. A number of potentially viable replacements exist for planar integrated optical circuits at this wavelength, namely Silicon on Sapphire [25], Silicon Germanium [26], Indium Phosphide [27], Aluminium Gallium Arsenide [28], Lithium Niobate [29], and Chalcogenide Glass (ChG) [30].

In addition to the obvious requirements for compact low loss circuitry there is a further requirement related to interfacing to the pupil plane sampling in the telescope. This is most easily achieved with microlens arrays having customized spatial distributions, that typically have a Numerical Aperture (NA) ~0.2 and below. Efficient coupling with this sort of NA requires large waveguides with low index contrast for them to be single mode. On the interferometer chip, however, reasonably large NA single mode waveguides are required in order to implement the <0.5 mm bend radii needed to make the circuits compact. Thus there is a fundamental mismatch in core sizes and index differences between the sampling and processing that has to be accommodated through a mode transformer of some type. The transformer also has to be intrinsically very low loss as telescope photons are scarce and scattered light from mismatched modes cannot be tolerated in the interferometer as it may produce spurious interference.

Chalcogenide glasses (ChGs) offer a more flexible alternative in finding a solution to this mismatch. Laser writing [31] of chalcogenides for example provides a low index change which eases coupling difficulties. Unfortunately this significantly limits bending radii which need to be quite small to minimize circuit size to enhance stability and losses. Chalcogenide glasses also offer entire families of glass composition that can be tuned to generate a very wide range of index contrasts. In addition, simple vertical tapering methods have been demonstrated in ChG with extremely low transition losses that can be used to adapt the mode size from a small high index contrast waveguide that can be sharply bent to one which is optimized for coupling with lower NA microlens arrays [32]. This design flexibility and the future scalability to N band was the factor that motivated this design study and ongoing work to implement the final design as a functional device.

### 3. Waveguide design and low NA mode matching

Initially the waveguide design follows from [30] and uses the combination of Germanium, Arsenic and Selenium ($Ge_{11.5}As_{24}Se_{64.5}$) as the core material and a Sulfur substituted analog ($Ge_{11.5}As_{24}S_{64.5}$) as the under- and over-cladding. These materials were chosen as they can be deposited by thermal evaporation in a bulk like state and so require no annealing [33] which would introduce birefringence through the resulting structural rearrangements. The refractive index of these glasses were measured by spectroscopic elipsometry as 2.609 and 2.279 respectively [30] at 4 μm.

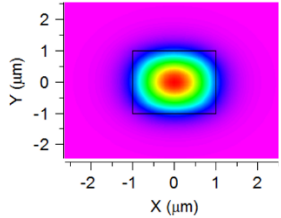

Fig. 1. The mode distribution of a single mode 2 by 2 μm waveguide of design as per text at 4 μm wavelength.

The maximum width and height for a single mode waveguide was determined by modelling using RSoft FEMSim in full vector mode for a square core buried in cladding glass. The resulting mode intensity profile for a 2 by 2 μm waveguide single moded down to 3.8 μm is shown in Fig. 1 for the TE (and equivalently due to symmetry also TM) propagating mode. The $1/e^2$ mode field diameter was 2.6 μm, matching the central lobe of an Airy disc formed in air with a numerical aperture of 0.79, or a Gaussian with minimal truncation formed with an NA of 0.82.

To attain high efficiency coupling to low NA systems, a 3-D taper with simultaneous width and height changes, or a pair of 2-D tapers can be used. The 2-D taper pair, whilst conceptually inferior, is considered as it eases mechanical alignment tolerances imposed by the fabrication methodology for the vertical taper [32]. The challenge is to keep the light confined in the fundamental mode of the now multimode waveguide, by designing to avoid mode coupling. Both device types were modelled with RSoft BEAMProp in full vector mode.

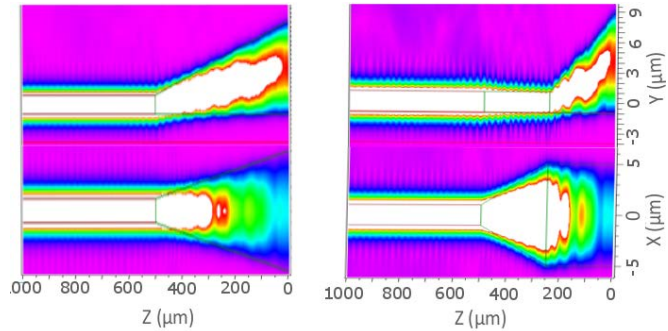

Fig. 2. The full 3-D taper simulation (left): Dual 2-D taper simulation; (right): Height reduces from 10 to 2 μm in first taper and width in the second. Beam propogation is from right to left.

Figure 2 shows simulations of a full 3-D and dual stage (vertical then lateral tapers) transforming a 10 by 10 μm waveguide fundamental mode to a 2 by 2 μm waveguide mode. It is clear from Fig. 2 that such devices work, but the transformation losses also need to be clarified.

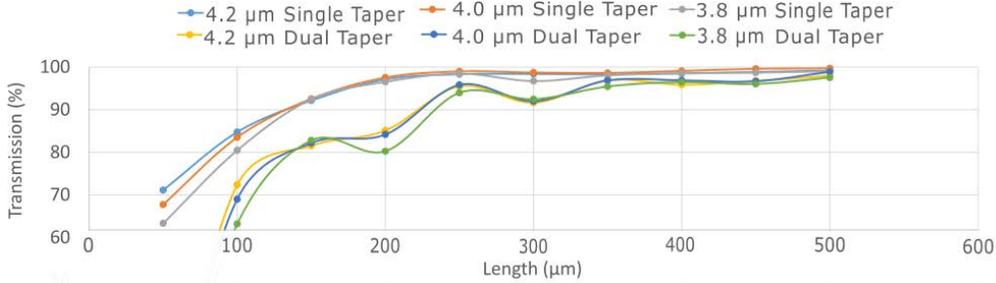

Fig. 3. Transmission for the two taper designs vs wavelength and total length.

The calculated transmission across the wavelength window of interest is displayed in Fig. 3 as a function of taper length (sum of two tapers for the dual case), treating all light not in the fundamental mode of a 2 by 2 μm waveguide as lost.

Figure 3 shows that the single taper is better behaved (the monotonic increase indicating nelegible mode coupling) creating a device 2 times shorter than the two taper system for the same throughput of ~95%. For taper lengths >250 μm, the 3-D device is essentially lossless. The main issue is in the fabrication tolerances, which requires the lithographic mask for the waveguide width definition to be aligned to the mechanically defined vertical taper to a tolerance <10 μm. This is a non-trivial exercise as the taper deposition mask is usually mechanically machined and so there are large tolerances at play and deposited fiducials are necessarily large and of low edge sharpness due to the tapering.

The fall back option is the dual taper system where the vertical tapering is separated from the width tapering resulting in significantly relaxed lithographic alignment tolerances. However as can be seen in Fig. 3 there is non-monotonic behaviour occuring as a function of length which is a clear indicator of mode coupling occuring. However the magnitude of this is acceptably small resulting in only slightly lower transmission (96% vs 99.5%) than the full 3-D case by a total length of 400 μm. Thus the dual taper is also quite acceptable, requiring more space on the chip for the benefit of simpler fabrication.

4. **The multimode interference coupler design and tolerances**

The most basic component in an interferometer is the combiner/splitter. The port counts and coupling ratios vary depending on the particular nulling architecture, with the simplest type being the Bracewell Interferometer [11] that requires a single 2x2 splitter with 50% coupling ratios [34]. Once the Bracewell Interferometer has provided a proof of concept, the next step is to use the architecture proposed by Angel and Wolf [35] that is referred to as the nulling interferometer in this paper and requires three of the same couplers.

Multimode Interference couplers (MMIs) are proposed for MIR application rather than evanescent couplers. MMIs have long been known to offer improved fabrication tolerances compared to most types of evanescent waveguide directional couplers [36] in addition to typically larger non-critical feature sizes. One factor in the improved tolerance of MMIs for example is the exponential dependence of coupling strength with waveguide separation in evanescent couplers. Such a level of sensitivity is not directly compatible with wide fabrication tolerances. The main figure of merit for couplers is the uniformity of the power split (referred to as the coupling imbalance) which is directly related to the null depth in a nulling interferometer.

Prior analyses [37] have not investigated the tolerances to all known process variables, especially for the MIR, and this is now considered in detail.

### 4.1 MMI design optimization

The requirement for a Bracewell Interferometer is that the MMI splits light equally between the two output ports. Using two input ports also means that a restricted interference class of MMI, as defined by Soldano [38], allows reduced device lengths. The use of tapered inputs, analyzed and optimized by Hill [39], enables improved excess loss performance. Accordingly, the length of the MMI is found approximately using Eq. (1):

$$L = \frac{M}{N} \frac{4nW^2}{3\lambda} \qquad (1)$$

where W is the width of the MMI, λ is the wavelength, n is the refractive index of the core material, N being the number of desired localized intensity maxima in the cross section (in this case 2) and M being the occurrence number of the N maxima along the device length (set to 1 to minimise the length) with M and N having no common factors [38,40–42].

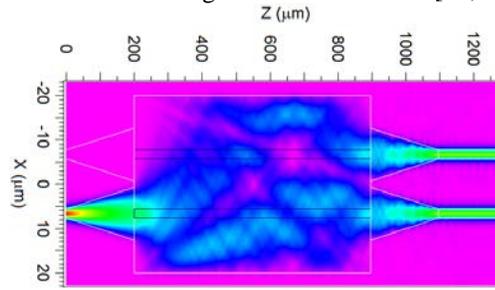

Fig. 4. An RSoft BEAMProp simulation of an MMI at 4 μm for an optimized design. Power is from the left and propagates to the right. The black lines in the middle of the MMI cavity are only to track the progress of the input light.

Equation (1) describes a restricted MMI that requires the input waveguides to be at positions 1/6[th] the MMI width, as shown in Fig. 4. The non-restricted equivalent is similar but with a length three times greater.

The splitter will cause a 90° phase shift between the output waveguides at a tolerance described in [39]. A basic nulling interferometer is thus formed by running the MMI, from Fig. 4, in reverse (right to left) to provide one output port with ~100% of the light and a null in the other.

The width of the MMI has a square relation to the MMI length and is the only free variable to be selected in Eq. (1). More exact modelling was undertaken with BEAMProp in full vector mode with a transverse grid size of 100 nm and a propagation step length of 500 nm. A full dispersion relation for the core and cladding refractive indices was incorporated into the model. For simulations that used the same dimensions, wavelength and refractive index the input mode was invariant and thus the 2 by 2 μm waveguide mode, computed by FEMSim, was used as the launch and monitor mode; otherwise the launch mode was computed directly by BEAMProp and also used as the monitor for the output powers.

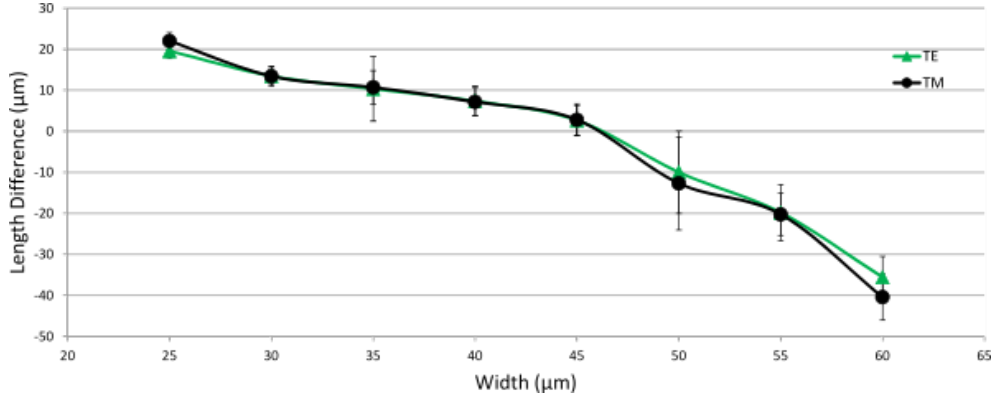

Fig. 5. The difference in the length for minimum imbalance subtracted with the length for optimum transmission vs MMI width. The error bars are from the numerical noise in the simulation.

Simulations indicated that the MMI length for a given width can be independently optimized in terms of equalization of the coupling ratios or the total throughput of the device [36]. Figure 5 shows the difference between these optimum lengths as a function of MMI width from BEAMProp simulations over MMI widths from 25 μm to 55 μm. The imbalance is found using Eq. (2), with bar port defined as that directly opposite to the input taper and the cross port being that diagonally opposite the input respectively. The imbalance defines a deviation from perfect 50/50 splitting (an imbalance of 0) and is used here in conjunction with the maximum transmission for device optimization.

$$\text{Imbalance } (\%) = \frac{\text{Bar port} - \text{Cross port}}{\text{Bar port} + \text{Cross port}} \times 100 \quad (2)$$

From Fig. 5 a width of 45 μm has an almost equal length for zero imbalance and optimum transmission, to within numerical simulation error, when averaged across TE and TM modes. Whilst Fig. 5 explicitly indicates that there is an optimum choice in the 45-50 μm width range: 45 μm was chosen as this provides the shortest MMI length. A device width of 45 μm (and corresponding length of 875 μm where the imbalance is zero for the TE mode as shown later in Fig. 6) was therefore selected as the MMI width for all further simulations.

*4.2 MMI width and length dependence*

Having an optimized design for the base MMI, its tolerance to fabrication errors needs to be elucidated. In terms of the MMI geometry, errors can occur independently in the length and width of the MMI during the mask fabrication step according to the accuracy of the mask writer, termed its critical dimension (CD) tolerance, the level to which the fabricated feature matches the design feature. During actual device fabrication, the length and width will be affected together by both the lithography and etching processes, and independently to any mask errors therefore adding cumulatively. During lithography, the initial imaging of the device onto the wafer, the final size of the printed feature compared to the mask feature size after resist development will depend upon a number of factors, (e.g. exposure dose, temperature humidity, develop time, etc). Similarly, during dry plasma etching, further size changes can occur depending upon the etch time, the degree of undercut or mask erosion, etc. Both processes affect all features in the same way causing the outline of the device to grow or shrink symmetrically around the nominal dimensions. Thus a more detailed length scan around the 45 μm design width, at 4 μm wavelength, is shown in Fig. 6.Fig.

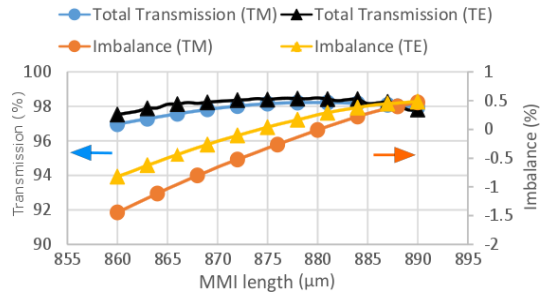

Fig. 6. Transmission and imbalance vs length for a 45 μm width MMI.

From Fig. 6 the transmission is essentially invariant over a ± 5 μm length variance for both TE and TM where the total light lost is ~1.5 % and ~1.9 % respectively. In this respect the MMI is robust against changes in its length. The length change also creates a shift in the phase, discussed more in [39]. The imbalance however is clearly not invariant around the optimum length of 875 μm, and it is the width to length relationship that is the main source of fabrication uncertainty in an MMI. A typical photomask writer achieves a CD tolerance of ± 0.1 μm at the feature sizes involved, this being a random error. This results in an imbalance, variation computed from Fig. 6, of ± 0.009 % for TE and ± 0.013 % for TM

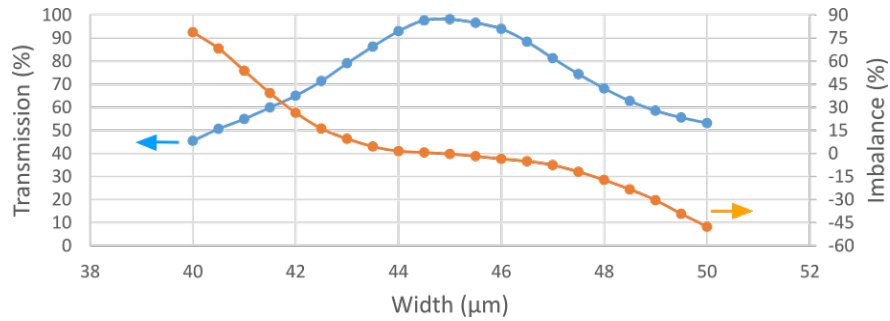

Fig. 7. Transmission and imbalance vs width for 875 μm length MMI.

Width is, from Eq. (1), a more sensitive parameter than length. For simplicity, from this point on, whilst both polarizations were modeled only the TM data is shown unless the TE showed a significant deviation. Figure 7 shows the transmission and imbalance as a function of width around the nominal design width. A CD tolerance of ± 0.1 μm produces a transmission variation of 0.35 % (not a concerning change) but an imbalance variation of ± 0.24 %, which is considerably more than for the length variation.

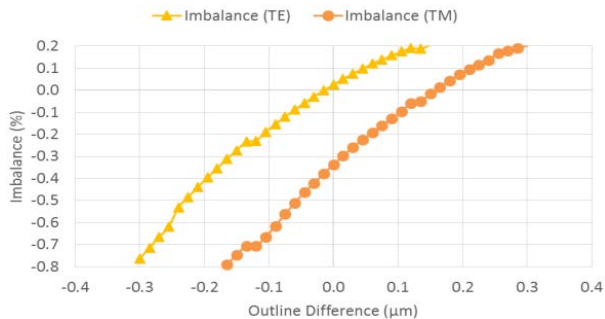

Fig. 8. The imbalance after the outside boundary of the MMI is varied for both TE and TM.

It is clear that the width has a greater effect than the length, however this is not the whole story. As noted above, lithography and etch biases result in a uniform error over the complete outline of every structure on the wafer. A realistic fabrication tolerance of ± 0.1 μm for every dimension of the MMI, including the tapers and gap between them, was simulated independently with the results shown in Fig. 8. Note the grid size was changed to 50 nm to allow better resolution of the device outline.

From Fig. 8 the imbalance has a similar dependence on the outline of the MMI to the width with an imbalance variation of ± 0.35 % per ± 100 nm outline change.

*4.3 Refractive index*

The refractive index of the MMI will remain constant throughout a single device or a small group of devices (at least over a device <1 mm long), however there will be intra- and inter-wafer variations in the refractive indices of the core and cladding. Experience with the films involved over many depositions indicates a total variation of 0.01 in refractive index is the worst case. Accordingly, the performance of the optimized design above was modelled over a wide range of core and cladding indices.

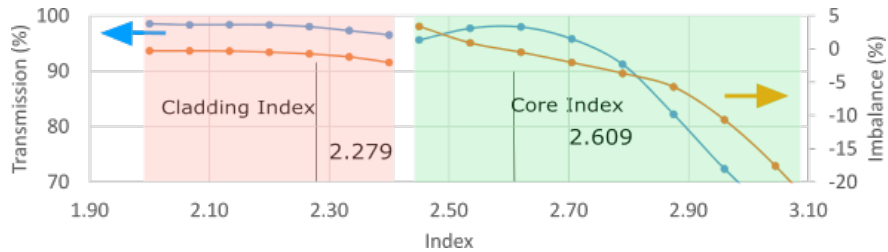

Fig. 9. Transmission and imbalance vs core (green shaded area) and cladding (red shaded area) refractive index.

Figure 9 shows the transmission and imbalance relation based on the refractive indexes, which where varied independently. The number indicated by the vertical line was the design refractive index for the optimized device and the value used when varying the other index. As would be expected, the performance is relatively insensitive to the cladding index exhibiting an imbalance variation of ± 0.0015 % per ± 0.005 change in cladding refractive index around the design point (2.279) and negligible change in the transmission. As suggested by Eq. (1), however, the performance is more sensitive to the core index. Around the design point (2.609), the imbalance change is ± 0.09 % per ± 0.005 of refractive index change. The transmission again is effectively unchanged.

*4.4 Layer thickness*

The other remaining device geometry variable is the film thickness. The expected thickness variation over a single device (millimeter scale) would be very small (sub nanometer) but over centimeter scales and between wafer runs more significant variations occur, typically a maximum range of ± 0.5 % or a total range of 20 nm for the design considered here.

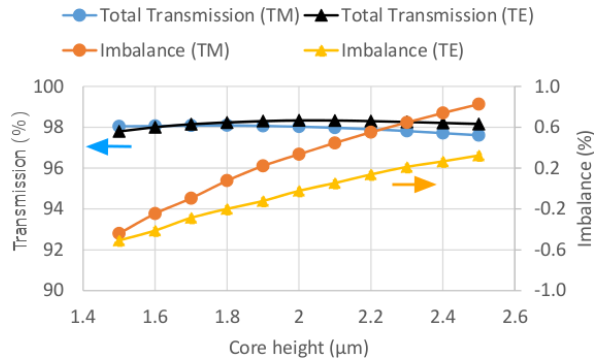

Fig. 10. Transmission and imbalance vs core thickness.

Figure 10 shows a simulation from BEAMProp of the effect of core thickness variation on the reference design. For the ± 10 nm achievable accuracy the change in imbalance around the reference value of 2.00 μm is ± 0.011 % for TE and ± 0.0010 % for TM.

### 4.5 Waveguide taper width

Whilst the impact of changes in the taper width is captured in the lithography and etching outline error considered above, the effects of solely varying the taper widths are considered here as it offers some further design flexibility and increased fabrication robustness for tapers a little narrower than optimum. The tapers increase robustness and reduce insertion losses and originated from research outlined in [39].

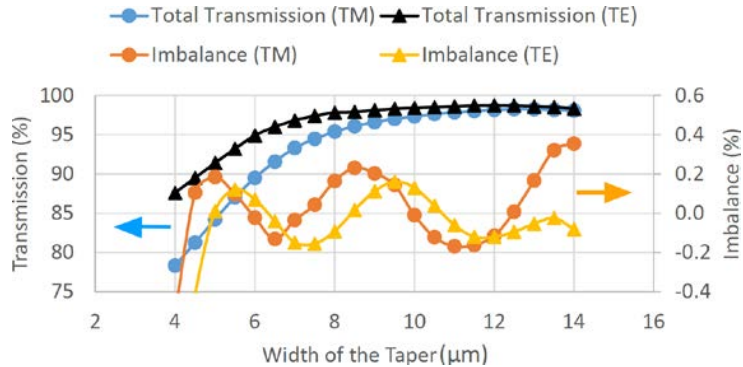

Fig. 11. Transmission and imbalance vs taper width.

The expected optimum width is calculated as 3/10th the MMI width, 13.5 μm for a 45 μm wide MMI. As is shown in Fig. 11, varying the width of both tapers simultaneously has minor effect on the transmission: decreasing the width from the "optimum" value of 13.5 μm there is only a gradual fall in the transmission until the decline accelerates for widths 8 μm and smaller. The taper length is chosen to be short enough to keep the light single moded by preventing mode coupling. In this simulation the length was set to 200 μm which is consistent with Fig. 3. As can also be seen in Fig. 11, there are in fact several alternate widths at which excellent imbalance can be obtained, and that the performance is not highly sensitive to the taper width.

### 4.6 Fabrication tolerance summary

The effects of the various possible sources of fabrication tolerance error on the imbalance are now summarized in Table 1.

Table 1. List of Fabrication sources, corresponding imbalance and area of impact on the full nulling device

| Source | ± Imbalance change (%) | Impact in nuller tree |
|---|---|---|
| **Mask error length (± 0.1μm)** | 0.013 | **Per MMI** |
| **Mask error width (± 0.1μm)** | 0.240 | **Per MMI** |
| **Outline Lith/etch bias (± 0.1μm)** | 0.350 | **All MMIs together** |
| **Core Index (± 0.005)** | 0.060 | **All MMIs together** |
| **Cladding Index (± 0.005)** | 0.002 | **All MMIs together** |
| **Core Thickness (± 0.01μm)** | 0.011 | **All MMIs together** |

From this it is clear the dominant error sources are the random mask width error, and the lithography and etching bias outline errors.

*4.7 Wavelength dependence*

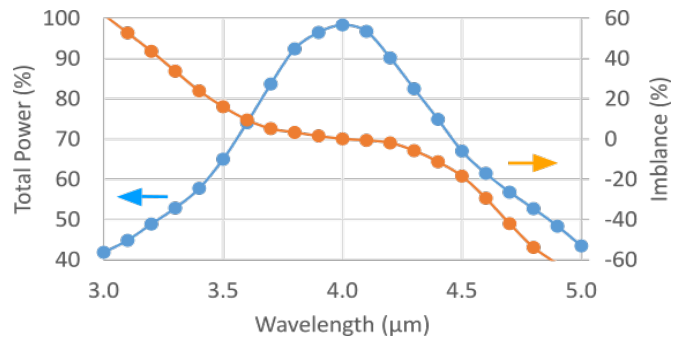

Fig. 12. Transmission and imbalance vs wavelengths for an optimized MMI.

Whilst the wavelength is not a fabrication variable it is nonetheless a quantity of operational interest when nulling performance is considered. The bandwidth of a MMI is expected to be wavelength dependent due to the $1/\lambda$ term in Eq. (1) for the 3 dB coupling length. The reference design was modelled using BEAMProp across a range of wavelength with the results displayed in Fig. 12.

The design of the MMI was optimized for a wavelength of 4 μm as is evident from the peak in the transmission in Fig. 12. The MMI transmission around the design wavelength of 4 μm, is 98 ± 6.5 % for ± 200 nm of optical bandwidth. The imbalance is minimized at the design wavelength as desired, and to a first order approximation, varies linearly with a proportionality constant of 0.0126 %/nm in the ± 200 nm operating bandwidth.

## 5. **Beam combiner**

The MMI has so far been presented on its own, however, three are required for the architecture proposed by Angel and Wolf [35] – which involved interfering two sets of two beams separately and then interfering the resulting output together.

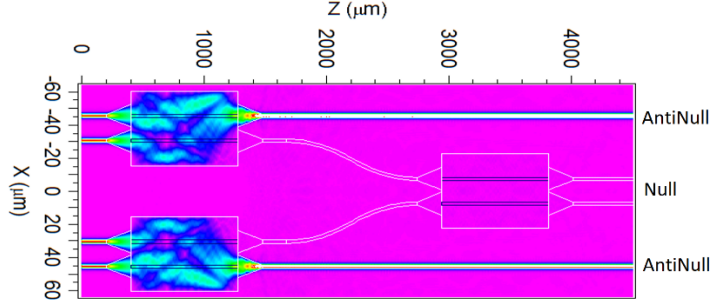

Fig. 13. Three MMIs in a 2:1 Angel and Wolf configuration [35]. The black lines in the centre of each MMI are only for simulation purposes they are in no way physical.

The design shown in Fig. 13 is as follows: The star light enters the first two MMIs and, by appropriate phase changes applied to the input light, is almost entirely directed to the outside (top and bottom in the diagram) output waveguides (termed the antinulls). To achieve this result the input wavefronts, to the two input MMIs, must have a 90º phase offset controllably imposed upon them. By tuning the phase offset, light from a close companion (incoherent with the star light), is directed into the (middle) third MMI. The middle MMI is a simple power combiner where no nulling is intended where an appropriate scan of the phase offset provides detailed information of surrounding light sources. This operation is herein termed single nulling due to a null of the host star only occurring at the first two MMIs and is similar to that used for the beam combination in *Dragonfly* [21]. A small proportion of the star light exits from each input MMI into the third MMI, and can be interfered again with appropriate imposed phase shifts to send most of it to one of the two central output waveguides. This is a different operational mode herein termed double nulling as per Angel and Woolf [35] and illustrated in Fig. 13 with the exception of one null port becoming an antinull port.

The figure of merit of a nulling interferometer is the Extinction ratio, which is defined as

$$\text{Extinction (dB)} = 10\text{Log}\left[\frac{\sum \text{Null}}{\sum \text{AntiNull}}\right], \quad (3)$$

where the Null is the power from the on-axis star light that has been suppressed by the architecture in Fig. 13 and the AntiNull is the power in the channels the star is directed into. In Fig. 13 the null channels are the middle two waveguides (directly after the third MMI) and the AntiNull channels are the highlighted channels at the top and bottom of the frame.

As shown in Table 1; the main sources of error are the, per MMI random mask width variation and the lithography and etch bias outline error. To evaluate the likely impact of these factors on the nulling performance over the operating bandwidth, a Monte Carlo simulation was performed where three random MMI widths are chosen within the error bounds ($\pm$ 100 nm) and then a random outline error applied to all three MMIs together within the error bounds ($\pm$ 100 nm). The chosen system then had the extinction ratio simulated across the operating window, and the process was repeated 100 times. To do this using BPM would however be unacceptably slow, and so the individual MMI performances were represented in matrix form with wavelength dependent parameters.

A series of targeted BPM simulations were used around the nominal design to fit the variation of the MMI coupling efficiency (k) and the transmission ($\alpha$) across wavelength and over the full range of width and outline variations. The coupling matrix in equation 4,

$$\alpha \begin{vmatrix} \sqrt{1-k} & i\sqrt{k} \\ i\sqrt{k} & \sqrt{1-k} \end{vmatrix}, \quad (4)$$

was then used for each MMI individually with the α and k over the wavelength range chosen from the interpolated table of values based on the random parameter choices (and $i^2 = -1$). The first two MMIs (as shown in Fig. 13) were set to have inputs of 1 and i to represent two incoming coherent wavefronts at a 90° phase difference. The middle MMI uses the outputs of the first two MMIs. All waveguide and bend losses have been excluded in this model as their impact would be minimal based on the short distance between the MMIs and any significant loss would be dependent on a range of different fabrication errors not discussed in this paper.

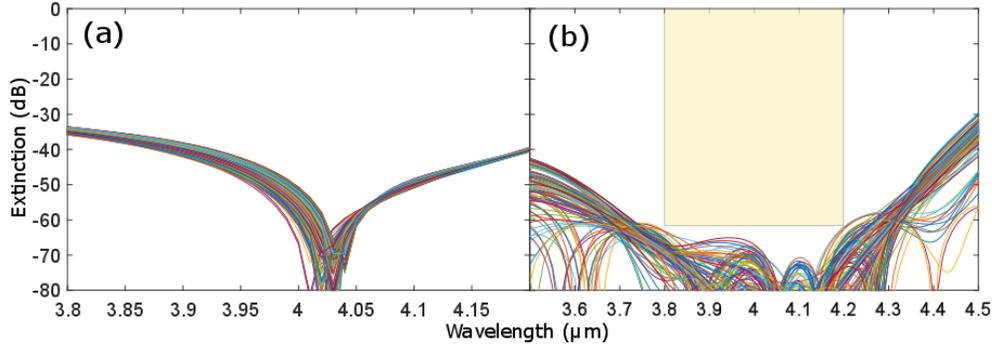

Fig. 14. Monte-Carlo simulation of extinction for three coupler MMI based Bracewell nuller over a range of realistic fabrication tolerances and wavelengths. (a) single nulling operation and (b) double nulling operation – the yellow box represents the bandwidth of 3.8 to 4.2 µm.

Figure 14 shows the results of the Monte Carlo simulation for 100 nullers dimensioned as described above for the TM mode extinction (worst case) of the nulling interferometer. Figure 14 (a) illustrates that a 40 dB extinction is achieved across a bandwidth of 3.9 to 4.2 µm when simulating a single nulling operation whereas (b) has a 60 dB null using a double null operation over the same bandwidth.

## 6. Conclusion

A beam combiner for a Bracewell inspired nulling interferometer has been designed and rigorously simulated to elucidate fabrication tolerances. It has been shown that the selected coupler, an MMI, is robust against changes in width and the total outline providing a nulling device with a reasonably achromatic response over the bandwidth of 3.9 to 4.2 µm.

For the operational bandwidth of 3.9 to 4.2 µm, a > 40 dB null is expected in single nulling mode. An even better extinction can be reached with a smaller bandwidth, exceeding 60dB for narrowband observations though at the expense of total power arriving at the detector. In double nulling mode, extinction exceeding 60 dB is available over the whole 3.8 to 4.2 µm operating bandwidth. At these null depths, scientifically meaningful observations are achievable with, for example, EGPs and exoplanets within 1 AU of the host star. These types of measurements work in tandem with established coronagraphic measurements that cannot penetrate as close to the host star and thus cannot resolve planets inside the habitable zone. Work is currently in progress to fabricate and characterize the optimized device designed here, and verify the predicted performance experimentally.

**Funding**

This research was supported by the Australian Research Council (ARC) Centre of Excellence for Ultrahigh bandwidth Devices for Optic Systems (CUDOS) project CE110001018.

**Acknowledgments**

Thank you to the Australian National Fabrication Facilities for their financial support for the RSoft design tools.